\documentclass{PoS}

\def\asec{\ifmmode ^{\prime\prime}\else$^{\prime\prime}$\fi}

\def\degs{\ifmmode ^{\circ}\else$^{\circ}$\fi}
\def\amin{\ifmmode ^{\prime}\else$^{\prime}$\fi}
\def\asec{\ifmmode ^{\prime\prime}\else$^{\prime\prime}$\fi}

\def\degs{\ifmmode ^{\circ}\else$^{\circ}$\fi}
\def\amin{\ifmmode ^{\prime}\else$^{\prime}$\fi}

\def\EE#1{\times 10^{#1}}

\def\cm{\mbox{\,cm}}

\def\cm3{\mbox{\,cm$^{-3}$}}

\def\ergshz{\mbox{~erg~s$^{-1}$~Hz$^{-1}$}}

\def\lsim{\!\!\!\phantom{\le}\smash{\buildrel{}\over
 {\lower2.5dd\hbox{$\buildrel{\lower2dd\hbox{$\displaystyle<$}}\over
                                 \sim$}}}\,\,}
\def\gsim{\!\!\!\phantom{\ge}\smash{\buildrel{}\over
{\lower2.5dd\hbox{$\buildrel{\lower2dd\hbox{$\displaystyle>$}}\over
                               \sim$}}}\,\,}

\title{Arp 299-A: More than ``just'' a prolific supernova factory}

\ShortTitle{Core collapse supernovae and starbursts}

\author{\speaker{Miguel A. P\'erez-Torres}\thanks{We acknowledge support from grant AYA2009-13036-C02-01, sponsored by the Spanish MICINN. M.A.P.T also acknowledges support from the Autonomic Government of Andalusia under grants P08-TIC-4075 and TIC-126. This research has also benefited from research funding from the European Community Framework Programme 7, Advanced Radio Astronomy in Europe, grant agreement no.: 227290. The European VLBI Network is a joint facility of European, Chinese, South African and other radio astronomy institutes funded by their national research councils.}, Antonio Alberdi, Cristina Romero-Ca\~nizales \\
        Instituto de Astrofísica de Andalucía - CSIC, 18080 Granada, Spain\\
        E-mail: \email{torres@iaa.es, antxon@iaa.es, cromero@iaa.es} 
               }

\author{Marco Bondi \\
Istituto di Radioastronomía - INAF, 40129 Bologna, Italy\\
E-mail: \email{bondi@ira.inaf.it} 
}

\author{Antonis Polatidis
\\
ASTRON, 7990 AA Dwingeloo, The Netherlands\\    
E-mail: \email{polatidis@astron.nl}
}

               \abstract{
We present partial results from our monitoring of the nuclear region of the starburst galaxy IC 694 (=Arp 299-A)
at radio wavelengths, aimed at discovering recently exploded CCSNe, 
as well as to determine their rate of explosion,  which carries crucial 
information on star formation rates and starburst scenarios at work.

Two epochs of eEVN observations at 5.0 GHz, taken in 2008, revealed the presence of a rich cluster of compact radio emitting sources in the central 150 pc of  the nuclear starburst in Arp 299A. The large brightness temperatures observed for the compact sources indicate a non-thermal origin for the observed radio emission, implying that most, if not all, of those sources were young radio supernovae (RSNe) and supernova remnants (SNRs). 
More recently, contemporaneous EVN observations at 1.7 and 5.0 GHz taken in 2009 have allowed us to shed light on the compact radio emission of the
  parsec-scale structure in the nucleus  of Arp 299-A. Namely, our EVN observations have shown that one of the compact VLBI sources, A1,
  previously detected at 5.0 GHz, has a flat spectrum between 1.7 and
  5.0 GHz and is the brightest source at both frequencies. The morphology, radio luminosity, spectral index and ratio of
  radio-to-X-ray emission of the A1-A5 region allowed us to identify A1-A5 with  long-sought AGN in
  Arp 299-A. This finding may suggest that both starburst and AGN are 
frequently associated phenomena in mergers. 
Finally, we also note that component A0, identified as a
young RSN, exploded at the  mere distance
of two parsecs from the putative AGN in Arp 299-A, which
makes this supernova one of the closest to a central supermassive
black hole ever detected. 
               }

\FullConference{10th European VLBI Network Symposium and EVN Users Meeting: VLBI and the new generation of radio arrays, September 20-24, 2010, Manchester UK}

\begin{document}

\section{The need for high-angular resolution in LIRG studies}

A large fraction of the massive star-formation at both low- and
high-$z$ has taken place in (U)LIRGs. Thus, their implied high
star-formation rates (SFRs) are expected to result in CCSN rates a
couple of orders of magnitude higher than in normal
galaxies. Therefore, a powerful tracer for starburst activity in
(U)LIRGs is the detection of CCSNe, since the SFR is directly related
to the CCSN rate.  However, most SNe occurring in ULIRGs are optically
obscured by large amounts of dust in their nuclear starburst
environments, and have therefore remained undiscovered by (optical) SN
searches.  Fortunately, it is possible to discover these CCSNe through
high-resolution radio observations, as radio emission is free from
extinction effects.  Furthermore, CCSNe are expected, as opposed to
thermonuclear SNe, to become strong radio emitters when the SN ejecta
interact with the circumstellar medium (CSM) that was ejected by the
progenitor star before its explosion as a supernova.  Therefore, if
(U)LIRGs are starburst-dominated, bright radio SNe are expected to
occur and, given its compactness and characteristic radio behaviour,
can be pinpointed with high-angular resolution, high-sensitivity radio
observations (e.g., SN 2000ft in NGC 7469 \cite{colina01}; SN 2004ip in
IRAS 18293-3413, \cite{pereztorres07}; SN 2008cs in IRAS 17138-1017,
\cite{pereztorres08a}, \cite{kankare08}; supernovae in Arp 299 \cite{neff04,pereztorres09,ulvestad09,pereztorres10},
Arp 220 \cite{smith98,lonsdale06,parra07}, Mrk 273
\cite{bondi05}). However, since (U)LIRGs are likely to have an AGN
contribution, it is mandatory the use of high-sensitivity,
high-resolution radio observations to disentangle the nuclear and
stellar (mainly from young SNe) contributions to the radio emission,
thus probing the mechanisms responsible for the heating of the dust in
their (circum-)nuclear
regions.

\section{e-EVN imaging of Arp 299-A}

Arp~299 (the merging system formed by IC 694 and NGC 3690) is the
``original'' starburst galaxy (Gehrz et al. 1983) and an obvious
merger system that has been studied extensively at many wavelengths.
An active starburst in Arp~299 is indicated by the high frequency of
recent optically discovered supernovae in the outer regions of the
galaxy.  Since the far infrared luminosity of Arp 299 is $ L_{IR}
\approx 6.5\times 10^{11} L_{\odot} $, the implied CCSN rate is of
$\approx$1.7 SN/yr. Given that 50\% of its total infrared emission
comes from source A (see Fig. \ref{fig,e-EVN}), it is expected that roughly 1 SN/yr will explode
in region A. Therefore, this region is the one that shows most
promises for finding new supernovae. Indeed, Neff et al. (2004) found
a new component in this region, by comparing VLBA observations carried
out in April 2002 and February 2003. 

We proposed e-EVN observations aimed at detecting the radio emission from recently exploded SNe in Arp 299.  Our observations, carried out in April 2008 and December 2008 at 5.0 GHz, resulted in the deepest images ever of Arp 299-A (see Figure \ref{fig,e-EVN}).  We found 26 compact components above 5$\sigma$ rms noise, whose nature can be only explained if they are SNe and/or SNRs, likely embedded in super star clusters, and may challenge the standard scenario
 that directly links far-infrared luminosity to a CCSN and star
 formation rate, since the apparent rate of CCSNe might be much higher than expected.
We leave, however, a detail discussion of this and other issues for future publications.

\begin{figure}[htbp!]
\centering
\includegraphics[width=0.8\textwidth]{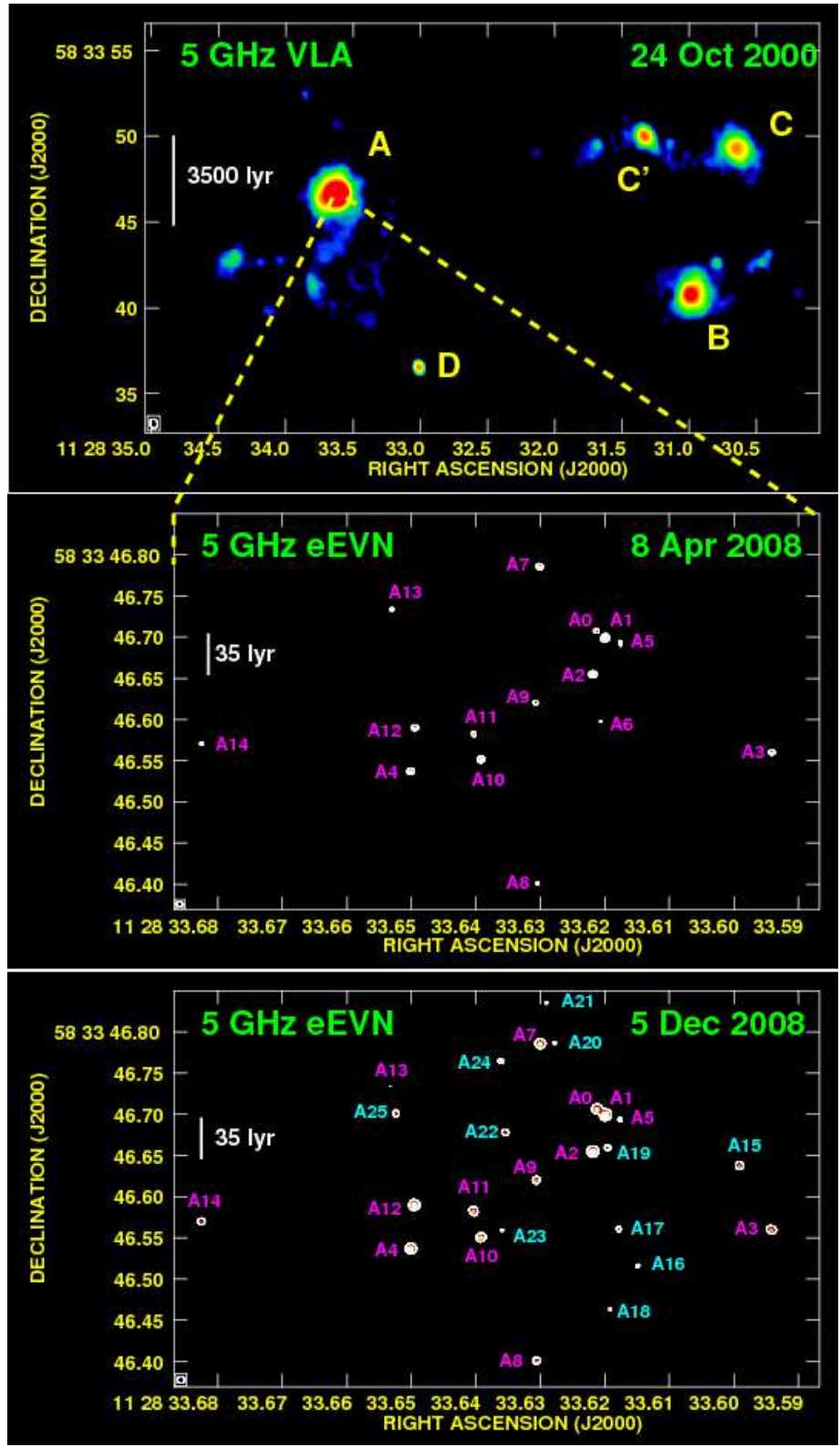}
\caption{ {\it Top panel:} 5 GHz VLA archival observations of Arp 299
  on 24 October 2000, displaying the five brightest knots of radio
  emission in this merging galaxy.  {\it Middle and bottom panels:} 5
  GHz e-EVN observations of the central 500 light years of Arp 299-A on
  8 April 2008 and 5 December 2008.  The off-source root-mean-square
  (r.m.s.) noise level is 39 $\mu$Jy/beam and 25 $\mu$Jy/beam for the
  middle and bottom panels, respectively, and show the existence of 15
  and 26 compact components with a signal-to-noise ratio (s.n.r.)
  equal or larger than five on 8 April 2008 and 5 December 2008,
  respectively.  The size of the FWHM synthesized interferometric beam
  was of (0.6 arcsec $\times$ 0.4 arcsec) for the VLA observations,
  and of (7.3 milliarcsec $\times$ 6.3 milliarcsec) and (8.6
  milliarcsec $\times$ 8.4 milliarcsec) for the e-EVN observations on 8
  April 2008 and 5 December 2008, respectively.}
\label{fig,e-EVN}
\end{figure}

\begin{figure}
\centering  
\includegraphics[width=90mm,angle=0]{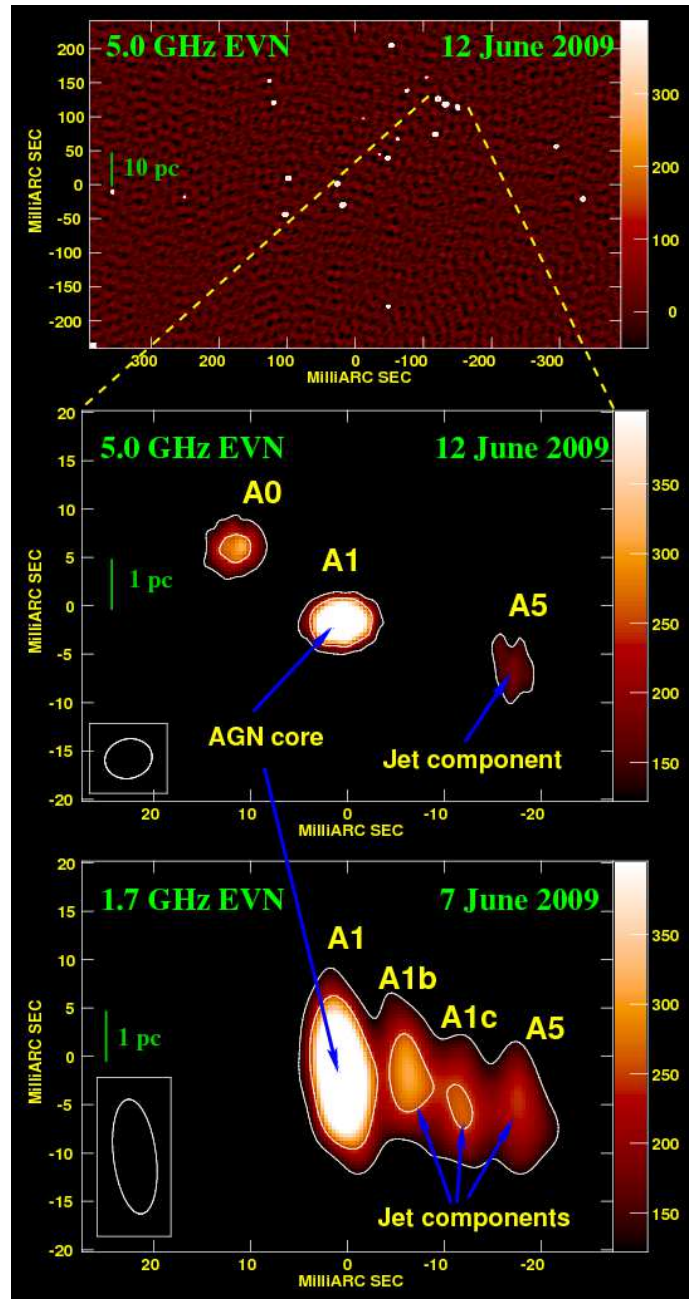} \\
\caption{ {\it Top:} 5.0 GHz full EVN image of the central 150 parsec
  region of the luminous infrared galaxy Arp 299-A (=IC 694), 
displaying a large number of bright, compact, nonthermal
  emitting sources, mostly identified with young RSNe and
  SNRs. The image center is at RA 11:28:33.63686 and
  DEC 58:33:46.5806.  {\it Middle and bottom:} Blow-ups of the inner
  8 parsec of the nuclear region of Arp 299-A, as imaged with the full
  EVN at 1.7 and 5.0 GHz.  The image center is at RA 11:28:33.61984
  and DEC 58:33:46.7006 in both panels. The morphology, spectral
  index and luminosity of the A1-A5 region are very suggestive of a
  core-jet structure.  The color scale goes from -50 $\mu$Jy/b up to 400
  $\mu$Jy/b in the top panel and from 125 $\mu$Jy/b to 400 $\mu$Jy/b in
  the middle and bottom panels. Contours are drawn at 5 and 10 times
  the off-source r.m.s. noise.}
 \label{fig,arp299a}
\end{figure}

\section{Summary and discussion}

VLBI observations of nearby CCSNe have allowed for a better
understanding of the physics, namely the determination of the
deceleration parameter, interaction between ejecta and presupernova
wind, characterization of the mass loss history and --sometimes-- the
explosion scenario, estimation of magnetic field and energy budget in
fields and particles. However, this wealth of information has been
obtained only for those supernovae that are bright ($L_{\rm peak}
\gsim 1.5 \EE{27}$ \ergshz), long-lasting (radio lifetimes of a few
years at least) and close enough ($D < 20$ Mpc), so that VLBI
observations can adequately resolve and monitor their expansion.  If
we consider that normal galaxies have small CCSN rates ($\lsim 0.01$
SN/yr), and that most CCSNe (Type IIP and Type IIb) have radio peaks
of a few times $10^{26}$\ergshz at most, this explains why so few
radio supernovae have been observed with VLBI in the last 20 yrs,
despite the VLBI arrays increasing their sensitivity.  One importan
way in which e-VLBI may contribute significantly in this field is in
the prompt response that it offers. For example, Type Ib/c SNe,
recently linked to long GRBs, are known to be rapidly evolving
($t_{\rm peak} \approx 10-20$ days) radio supernovae. Currently VLBI
arrays do not offer the needed dynamic scheduling to, e.g., react on a
nearby event, while e-VLBI offers such flexibility and, thanks to its
ability to carry out real-time correlation, allows for a potential
follow-up of the most interesting targets.

The CCSN rate in (U)LIRGs is expected to be at least one or two
orders of magnitude larger than in normal galaxies \cite{condon92},
and hence detections of SNe in (U)LIRGs offer a promising way of
determining the current star formation rate in nearby galaxies.
However, the direct detection of CCSNe in the extreme ambient
densities of the central few hundred pc of (U)LIRGs is extremely
difficult, as the optical and IR emission of supernovae is severely
hampered by the huge amounts of dust present in those regions, and can
at best yield an upper limit to the true CCSN rate.
Fortunately, it is possible to directly probe the star forming
activity in the innermost regions of (U)LIRGs by means of high
angular resolution, high-sensitivity radio searches of CCSNe, as radio does not suffer
from dust obscuration.
Current VLBI (and e-VLBI) arrays are starting to yield astonishing
results, thanks to their few-$\mu$Jy sensitivity and milliarcsec
resolution at cm-wavelengths. In particular, the findings in Arp 220 using global VLBI
 and Arp 299 (using the e-EVN), may challenge the standard scenario
 that directly links far-infrared luminosity to a CCSN and star
 formation rate, which is of much relevance for studies of starburst
 galaxies at high-z.

\begin{small}
\paragraph*{Acknowledgments}
The author acknowledges support by the Spanish
\textsl{Ministerio de Educación y Ciencia (MEC)} through grant AYA
2006-14986-C02-01,  and also by the \textsl{Consejería de Innovación, Ciencia y
Empresa} of  \textsl{Junta de Andaluc\'{\i}a} through grants FQM-1747 and
TIC-126. MAPT is a Ram\'on y Cajal Post Doctoral Research
Fellow funded by the MEC and the \textsl{Consejo Superior de
  Investigaciones Científicas (CSIC)}. 
 The European VLBI Network is a joint facility of European,
Chinese, South African and other radio astronomy institutes funded by
their national research councils.  
e-VLBI developments in Europe are supported by the EC DG-INFSO funded Communication Network Developments project 'EXPReS', Contract No. 02662.
\end{small}

\end{document}